\documentclass{article}
\usepackage{ijcai23}

\usepackage{times}
\usepackage{soul}
\usepackage{url}
\usepackage[hidelinks]{hyperref}
\usepackage[utf8]{inputenc}
\usepackage[small]{caption}
\usepackage{graphicx}
\usepackage{amsmath}
\usepackage{amsthm}
\usepackage{booktabs}
\usepackage{algorithm}
\usepackage{algorithmic}
\usepackage[switch]{lineno}

\DeclareMathOperator*{\argmin}{argmin}

\usepackage{latexsym}

\title{A Generalized Look at Federated Learning: Survey and Perspectives}

\author{
Taki Hasan Rafi$^1$\and
Faiza Anan Noor$^2$\and
Tahmid Hussain$^{3}$\and
Dong-Kyu Chae$^1$\And
Zhaohui Yang$^{4}$
 \\ 
\affiliations
$^{1}$Hanyang University, South Korea\\
$^2$Ahsanullah University of Science and Technology, Bangladesh\\
$^3$upay (UCB Fintech Company Limited), Bangladesh\\
$^4$Zhejiang University, China\\
\emails
\{takihr, dongkyu\}@hanyang.ac.kr,
faizanoor.cse@aust.edu,
tahmid.hussain@upaybd.com,
yang\_zhaohui@zju.edu.cn
}

\begin{document}

\maketitle

\begin{abstract}
\textbf{Federated learning (FL)} refers to a distributed machine learning framework involving learning from several decentralized edge clients without sharing local dataset. This distributed strategy prevents data leakage and enables on-device training as it updates the global model based on the local model updates. Despite offering several advantages, including data privacy and scalability, FL poses challenges such as statistical and system heterogeneity of data in federated networks, communication bottlenecks, privacy and security issues. This survey contains a systematic summarization of previous work, studies, and experiments on FL and presents a list of possibilities for FL across a range of applications and use cases. Other than that,  various challenges of implementing FL and promising directions revolving around the corresponding challenges are provided. 
\end{abstract}

\section{Introduction}

{\it{Federated learning (FL)}} is a distributed machine learning strategy in which learning algorithms can be trained with the participation of multiple entities without having to share local data to a central location (server or data center). Both centralized (server-controlled) and decentralized (server-less) settings can be leveraged for this training scheme. Due to its interdisciplinary nature, FL has been a prominent focus of modern machine learning research. It has attracted researchers from a wide range of disciplines, such as security, privacy, distributed systems and machine learning. FL has demonstrated the potential in solving problems from multiple domains such as healthcare \cite{huang2019patient,li2020multi,xiong2020facing,chen2020federated}, edge computing \cite{wang2019adaptive,nishio2019client,he2020group}, finance \cite{zheng2021federated,cheng2021secureboost}, and recommendation systems \cite{ammad2019federated,tan2020federated}.




\textbf{Why is Federated Learning necessary?} In recent years, there has been an explosion in data generation as a result of the widespread adoption of personal gadgets and Internet of Things (IoT) devices. As a result, data and computational resources are frequently spread across  personal devices, different geographic locations, or institutions. The security and privacy of end users or organizations are seriously put at risk when data from different sources are combined and sent to a centralized server or shared directly between different organizations or data centers. Traditional machine learning approaches face serious challenges in this aspect as they usually rely on aggregating the raw data dispersed across various devices or institutions to a single central location (server or data center) for training models \cite{zhu2021federated}. Besides, traditional centralized machine learning methods suffer from other limitations as well such as high computational power requirements, longer training time, etc. On the other hand, FL not only aims to ensure security and privacy for the distributed raw data but also intends to provide solutions to other issues that traditional machine learning systems face by utilizing the techniques employed in typically distributed machine learning approaches. As a result, FL emerged as a potential solution to the challenges faced by centralized machine learning approaches, especially in learning tasks where training data is decentralized in nature. 

With the aim of ensuring data privacy, FL prohibits the exchange of distributed raw data among the collaborating entities. It only allows the exchange of the intermediate data among the participating entities \cite{yang2020energy,chen2020joint}, which in most cases are the updated local gradients or models from the clients. As a result, FL brings the code to the data, whereas traditional machine learning approaches do the opposite, i.e., bringing the data to the code ~\cite{liu2021blockchain}. FL adopts this principle to eliminate the risk of sensitive personal information leakage. Besides, immediate aggregation of the updates (gradients or models) is performed as early as possible to add an extra layer of security against sensitive information leakage ~\cite{kairouz2021advances}. Additionally, several privacy-preserving systems, e,g., differential privacy, secure multiparty computation, homomorphic encryption, etc. have been integrated into FL over the last few years in order to address potential attacks ~\cite{wang2019beyond}.

 \textbf{Contributions.} There have been some noteworthy survey works on FL. Survey written by \cite{yang2019federated} focused on proposing an extensive and secure FL framework where participating entities are mostly a small number of enterprise data owners. \cite{kairouz2021advances} reviewed different aspects of FL, with a particular emphasis on cross-device FL. \cite{li2020multi} summarized previous works on FL from a systems perspective in addition to presenting a thorough taxonomy of FL. Among the recent works, the survey by \cite{li2021survey} is a noteworthy one that also focuses on FL system architecture and design. Some studies examine only a single aspect of FL, threats to FL \cite{lyu2020threats} or the application of FL in a particular field, e.g., FL specific to resource-constrained IoT devices \cite{imteaj2023federated}. Majority of early surveys review FL from a certain perspective such as system architecture, federation settings, taxonomy of different aspects related to FL system design, distributed training techniques, etc. So there was a lack of fully grounded survey paper in respect to covering all the major aspects of FL. 
 
 Recent survey works, such as \cite{zhang2021survey,aledhari2020federated,abdulrahman2020survey} have attempted to provide a broad perspective on the topic of federated learning. However, some of these surveys have failed to fully address certain areas; for instance, the applications of FL over a majority of fields are not well addressed. Some other surveys are primarily focused on the design and management of resources, rather than providing a general overview of the field. Compared to the existing surveys, our contributions can be structured in three folds:

 \begin{itemize}
     \item Our survey primarily provides a succinct yet comprehensive summary of the current state of FL, filling the gaps in existing surveys, with a focus on aligning the perspectives and addressing the need for a generalized overview of FL. 
     \item This survey also presents a summary of methods and algorithms used in FL, including privacy-preserving methods, popular FL algorithms, and their communication overhead and privacy protection differences.
     \item Additionally, it highlights system architectures, challenges and future research directions, as well as summarizing the key characteristics and recent empirical applications of FL.
 \end{itemize}

\textbf{Organizations.} To start with, we first present a general task definition with
the formulations of FL in Section 2. In Section 3, we briefly discuss about the downstream FL system architecture. Considering the core FL system based on different architectures, we summarily discuss the various methods considering multiple types of FL architectures in Section 4. In Sections 5 and 6, we put insights on potential applications of FL and diverse challenges in FL domain. 
In Section 7, we explore potential future directions. The final section offers our conclusion.

\section{Preliminaries: Federated Learning}

Generally, a FL system consists of two main entities,
i.e., the FL server and data owners. 

Let $N$ be the set of data owners, where $N = \{1,......., n\}$. Each of these owners possesses their own private dataset $D_{i}$. Using $D_{i}$, every data owner $i$ trains a local model $w_{i}$ and sends only the local model weights to the server. Afterwards, a global model $w_{G}$ is generated aggregating all collected local model weights $w_G = \frac 1 n\sum_{i=1}^n w_i$. This differs from the centralized training where training takes place centrally after aggregating data from individual sources. In a typical FL system, there always exist three phases:

\subsubsection{Phase 1 (Hyper-parameter Initialization):} In this phase, the server performs corresponding data requirements, task requirements, and specifies the hyper-parameters of the global model. Then, the server broadcasts the tasks and global model  $w^0_{G}$  to selected participants.

\subsubsection{Phase 2 (Local Model Training):} In this phase, the local model parameters of each are updated as they undergo training individually, using an individual participant's private data and model. Let
$i$ be a participant at iteration $t$. The goal is to find optimal model parameters $w^t_{i}$ that minimize the loss function $\mathcal{L}(D_i;w_i^t)$:
\begin{equation}
   w^{t*}_{i} = \argmin_{w^t_{i}} \mathcal{L}(D_i; w^t_{i})  
\end{equation}
The updated local model weights are thereafter sent to the server for aggregation.

\subsubsection{Phase 3 (Global Model Update):} 
Here the global model aggregation is done as the
server aggregates the local models from participants. Subsequently, the updated global model parameters  $w^{t+1}_{i}$  is sent back to the data owners.
The global loss function $\mathcal{L}(
D_i;w^t_{G})$ is minimized as:
\begin{equation}
\mathcal{L}(w^t_{G})=
\frac{1}{n}\sum_{i=1}^{n}\mathcal{L}(D_i;w^t_{i}) 
\end{equation}

\begin{figure}[t]
    \centerline{\includegraphics[width=8.5cm]{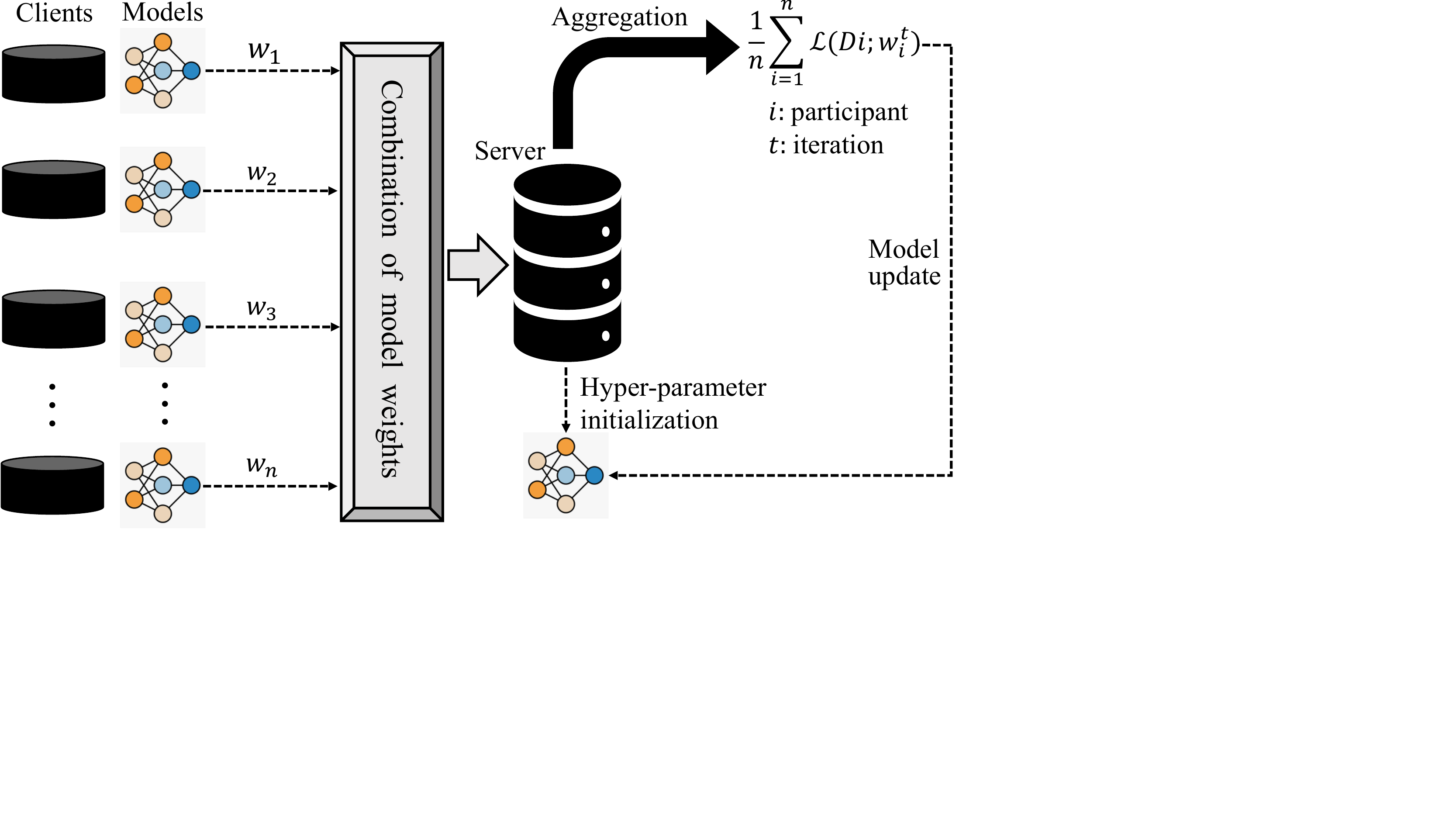}}
    \caption{Overview of FL.}
   \label{fig:fig2}
\end{figure}

As long as the global loss function does not converge, the steps 2-3 are repeated \cite{lim2020federated}. These are indispensable steps of FL and a classical algorithm for aggregation of local models is called FedAvg algorithm. Figure 1 depicts an overall intuition of FL.

\subsection{Categorization}

There are three categories of FL on the basis of distribution characteristics \cite{yang2019federated}. Let the data possessed by each data owner $i$ be symbolized as matrix $M_i$. Here, $X$ denote features and $Y$ denote label space. Let ID be sample space where each row indicates a sample and each column indicates a feature.

\begin{itemize}

\item \textbf{Horizontal Federated Learning (HFL)} is generally preferred by most FL practitioners. It is applicable to scenarios where that datasets differ in samples yet share the same feature space. HFL can be summarized as:

\begin{equation}
    X_i=X_j , Y_i = Y_j , ID_i \neq ID_j \quad \forall M_i, M_j, i \neq j 
\end{equation}

\item \textbf{Vertical Federated Learning (VFL)} applies to the scenarios where two data sets differ in feature space but have a common sample space. VFL can be summarized as:

\begin{equation}
    X_i \neq X_j , Y_i \neq Y_j, ID_i = ID_j \quad \forall M_i , M_j,i \neq j 
\end{equation}

\item \textbf{Federated Transfer Learning (FTL)} is applicable to the cases where along with differing samples, the two data sets also differ in feature space. FTL can be summarized as:
\begin{equation}
   X_i \neq X_j , Y_i \neq Y_j ,  ID_i \neq ID_j \quad \forall M_i , M_j,i \neq j
\end{equation}

\end{itemize}

\section{FL System Architecture}
Generally, the architecture of a FL system can be divided into 4 layers~\cite{liu2015survey}, which are: a) Presentation, b) User services, c) Federated training and d) Infrastructure.
To summarize, user interactions happen through the Presentation layer and the User services layer manages expected functionalities such as monitoring, log, steering, etc.

Distributed training of federated models occurs at the layer called Federated training layer, during the training phase following the execution of Federated Learning Execution Plan (FLEP). All physical resources such as storage resources, and computing resources, etc are managed at the Infrastructure Layer. Figure 2 illustrates the summary of FL system architectures. 

\textbf{Presentation Layer.} Interaction among users and Federated learning systems occurs through the presentation layer, which exist in the form of User Interfaces (UI). UI can be of two forms, one of which is a textual interface and the other is a graphical interface. Additionally, presentation layer shows the current status of the training process. FATE~\cite{Cai_2022} provides a graphical UI, which brings practicality and user-friendliness as the users can easily construct a Federated model by dragging and dropping. In contrast, textual interfaces construct  FL models using scripts or command line interface (CLI), and this is provided by frameworks such as TensorFlowFL, PaddleFL, PySyft, and FATE~\cite{Cai_2022}.

\textbf{User Services Layer.} User services layer monitors training process and real-time exectution status to check if the training is occuring normally using a log service. The log generated during the training process also participates in debugging and it adjusts the model accordingly.

FATE~\cite{Cai_2022} allows users to regulate the training process in case of unprecedented errors, through a graphical monitoring board. User services layer also helps in the understandability or interpretability of an FL system, which refers to the explanation of an FL system in a comprehendible way. This explainability  mainly explains the data representation within an FL model and results of training process. Providing interpretability and explainability concurrently remains an unsolved challenge even though Shapley values provided interpretability to a great extent. 

\textbf{Federated Training Layer.} The training procedure is carried out via the Federated Training Layer using distributed computation and data resources. The parallelization, scheduling, and fault-tolerance modules make up this layer. 3 kinds of parallelization exists (model, data, and pipeline).

A Scheduling Plan (SP) for executable jobs is created through the Federated scheduling module in order to fully utilize distributed computing resources and avoid bottlenecks in the training process. Then the updates are assembled to procure the final model. The obstructions in task execution are manipulated by the FL fault-tolerance mechanism. Techniques such as checkpoints, reboots, and task replications diminishes the aftermath of detected failures in this layer.

\textbf{Infrastructure Layer.} The interface between an FL system and distributed resources, such as computing storage, network, and data resources, is provided by the infrastructure layer. Three components make up this layer, such as a) distributed execution module, b) data transmission module, and c) data security module.

\begin{figure}[t]
    \centerline{\includegraphics[width=8.5cm]{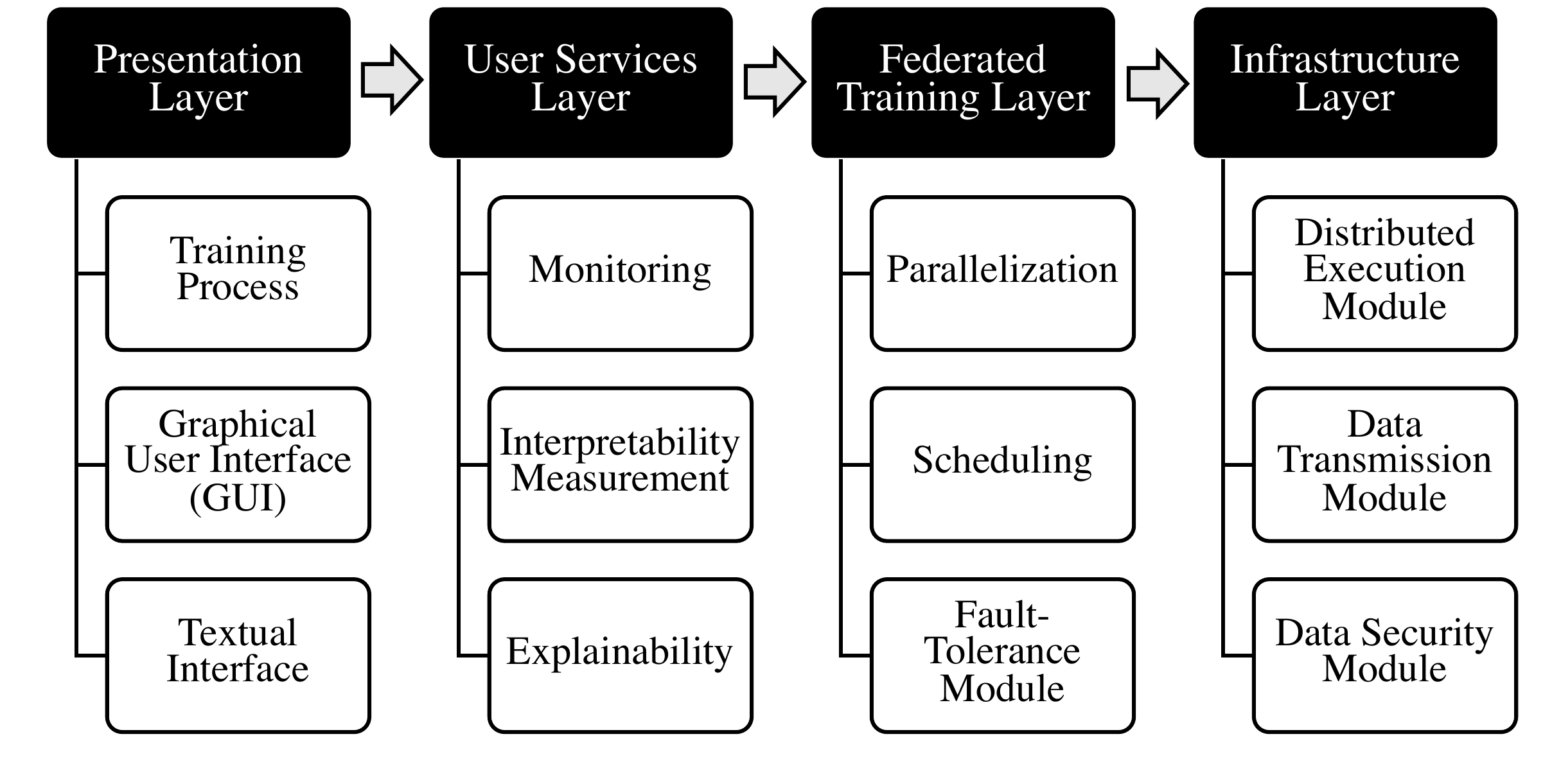}}
    \caption{FL system architectures.}
   \label{fig:fig2}
\end{figure}

The distributed execution module executes specific tasks using distributed computing resources and to safeguard the raw data used for training, the data security module often applies Differential Privacy (DP)  and encryption methods, such as homomorphic~\cite{hardy2017private}. While intermediary data such as weights or gradients can be exchanged between distributed computer resources, the raw data itself cannot be transported directly. The data transmission module leverages data compression techniques~\cite{bergou2022personalized} to enhance the effectiveness of data transport, alongside implementing Federated Learning Execution Plan (FLEP), created at the training layer.

\section{Insights on Methods}
Different algorithms or methods are leveraged by different practitioners to construct one or multiple types of FL architectures such as Linear Models (LM), Decision Trees (DT) and Neural Network (NN), etc. Apparently, most FL settings have horizontal data partitioning. Table \ref{tab:plain} summarizes several popular researches.

\begin{table*}[h!]
    \centering
    \footnotesize
    \begin{tabular}{p{4.6cm} p{2.4cm} p{2.0cm} p{2.5cm} p{2.2cm} p{1.5cm}}
        \toprule[2pt]
        References	& Name of Model(s) & 
        Implemented Model(s) & Communication Architecture & Data Partitioning & Privacy Mechanism\\
        \toprule

         ~\cite{bergou2022personalized}  & L2GD & LM & Centralized & Horizontal & -\\  \hline   
        ~\cite{makhija2022architecture}  & FedHeNN & NN & Centralized & Horizontal & -\\  \hline
~\cite{collins2021exploiting}  & FedRep & NN & Centralized & Horizontal & -\\  \hline

~\cite{li2019fair}   &SimFL & DT            & Decentralized                & Horizontal & Hashing \\ \hline
~\cite{mcmahan2017learning}  &FedAvg  & NN             & Centralized                & Horizontal & - \\ \hline
~\cite{hardy2017private}  &LR-FL  & LM            & Centralized                & Vertical & CM \\ \hline
~\cite{praneeth2019scaffold}  & SCAFFOLD  & LM, NN            & Centralized                & Vertical & CM \\ \hline

~\cite{marfoq2020throughput}  &DPASGD   & NN             & Decentralized                & Horizontal & - \\ \hline

~\cite{lin2020ensemble}  &FedDF  & NN             & Centralized                & Horizontal & - \\ \hline

~\cite{tan2020federated}  &FedRecSys   & LM, NN            & Centralized               & Horizontal & CM \\ \hline

~\cite{wang2021federated}  &FedMA  & NN             & Centralized                & Horizontal & - \\ \hline

~\cite{brendan2016communication}   &FedAvg  & DT, LM, NN             & Centralized     & Horizontal           &  CM, DP \\ \hline

~\cite{wang2020federated}  &SplitNN  & NN             & Centralized                & Vertical & - \\ \hline

~\cite{ammad2019federated}  & FedCF  & LM             & Centralized                & Horizontal & - \\ \hline

~\cite{liu2019boosting}    & FedForest  & DT           & Centralized          & Horizontal    & CM       \\ \hline

~\cite{mcmahan2017learning}    & FL-LSTM  & NN           & Centralized          & Horizontal    & -          \\ \hline

~\cite{hu2022oarf}   & OARF  & NN           & Centralized          & Horizontal    & DP, CM         \\ \hline

~\cite{liu2019boosting}    & FedXGB  & DT           & Centralized          & Horizontal    & CM       \\ \hline

~\cite{mohri2019agnostic}    & PRRR  & LM           & Centralized          & Horizontal    & CM        \\ \hline

~\cite{smith2017federated}   & Federated MT  & LM           & Centralized          & Horizontal    & -        \\ \bottomrule[2pt]
    \end{tabular}
    \caption{Comparison between existing works on FL. We denote Cryptographic Methods as CM, Differential Privacy as DP, Linear Models as LM, Neural Networks as NN, and Decision Tree as DT.}
    \label{tab:plain}
\end{table*}

~\cite{konevcny2016federated} updated the model weights based on distributed approximate newton algorithm (DANE) and stochastic variance reduced gradient (SVRG).~\cite{yurochkin2019bayesian} developed a Bayesian nonparametric framework to implement federated learning with multi-layer perceptrons. Their framework modeled local neural network weights presumably provided by each data server. They aggregated local models to procure a robust global network using a Beta-Bernoulli process informed matching process, and this approach surpassed the results of FedAvg \cite{mcmahan2017learning} for both IID and non-IID data partitioning.

~\cite{nikolaenko2013privacy} combined Yao's garbled circuits \cite{yao} and homomorphic encryption mechanisms in a horizontal FL Environment, and no information regarding input data was revealed, even though the model outputted the best fit curve. To maximize performance regarding privacy requirements,  each mechanism was implemented in different sections of the Ridge Regression algorithm. 

~\cite{zhao2018inprivate} implemented a gradient boosting decision tree (GBDT) and used differential privacy techniques to protect privacy for both the trained model as well as the individual data owners. In GBDT~\cite{chen2016xgboost}, regression trees from numerous data owners can be securely trained and combined. Each tree is trained locally without intervention among different parties.~\cite{cheng2021secureboost} proposed a novel system called \textit{SecureBoost}, a privacy-preserving lossless tree-boosting mechanism in a vertically partitioned federated setting, i.e. a federated data partitioning system with different feature sets yet common user samples. This system uses a privacy-preserving protocol for entity alignment first and then uses a highly efficient encryption mechanism to conduct boosting
trees among several parties. Similarly,~\cite{liu2019boosting} proposed a privacy-preserving lossless variation of the traditional random forest paradigm, called Federated Forest, which allows learning and training processes to be jointly conducted over clients or random forests across various regions in a vertical federated setting, where communicated data is encrypted. 

~\cite{wang2020federated} used the federated matched averaging (FedMA) algorithm with the goal to reduce the communication burden in federated systems that were implemented using multi-layer perceptron architectures like convolutional neural networks (CNN) and long short-term memory (LSTM). In this system, the global model was formed by averaging and matching information such as hidden states and channels for convolutional layers in a layer-wise manner, diminishing communication overhead to an extent. FedMA~\cite{wang2020federated} surpasses results of state-of-the-art CNN-based FL systems, FedAvg \cite{mcmahan2017learning} and FedProx \cite{li2020federated}.

To prevent sharing raw patient data and allow health entities to collaboratively train deep neural networks (DNNs),~\cite{vepakomma2018split} proposed a split learning method called SplitNN. This approach involves dividing a neural network into two parts, with each party participating in the training process only responsible for training a few layers of the network. The output from these layers is then passed to the party that has access to the labels, which completes the rest of the training process. Resource-constrained MEC (mobile edge computing) models were also constructed using horizontal data partition, linear models and neural networks.

\section{Applicability}

As FL has shown great promise to tackle the challenges that traditional machine learning approaches face, many industries have already begun to integrate FL into their work flows. This has resulted in many application areas of FL. This section provides an overview of the current mainstream application areas of FL.

\textbf{Healthcare.} One area that has been greatly impacted by the usage of federated learning is healthcare. Due to privacy laws and regulations, it is not possible for hospitals to just share data among themselves as the data contain very sensitive patient information. Besides, healthcare providers lack high-quality data due to reproducibility issues. However, the data contained in each hospital suffers from size limitations and cannot approximate real distributions.
As a result, using the data from a single healthcare facility to train machine learning models is challenging and does not always result in acceptable accuracy and generalization ability. 

FL has made it easier for hospitals to participate in learning tasks without compromising any privacy regulations.~\cite{li2020multi} proposed using FL for functional MRI (fMRI) data with an emphasis on the scarcity of high quality data in healthcare facilities. The authors in~\cite{huang2019patient} implemented CBFL (Community Based Federated Learning), an algorithm to predict patient hospital stay time and mortality. The algorithm was designed using EMR (Electronic Medical Records) data from eICU collaborative research database. 

FL has also been widely applied in medical imaging.~\cite{li2019privacy} proposed FL-based brain tumor segmentation using the BraTS 2018 dataset. FL has contributions in current pandemic situation too. An early diagnosis scheme for new COVID-19 cases has been developed via FL~\cite{ouyang2021novel}.~\cite{dayan2021federated} suggested a FL-based oxygen demand estimation model for patients with COVID-19 symptoms. Furthermore, drug discovery has also been one of the recent application areas of FL~\cite{xiong2020facing}.

\textbf{Edge Computing.} Since edge computing and FL are related in some aspects, this is one of the most well-known application areas of FL. The combination of edge computing and FL has made it possible for devices at the edge to train machine learning models without sending data to cloud sources.~\cite{wang2019edge} demonstrated that both Deep Reinforcement Learning (DRL) and FL can be used to effectively optimize mobile edge computing, caching, and communication. [Nishio and Yonetani, 2019] implemented federated averaging in practical mobile edge computing (MEC) frameworks to manage the resources of heterogeneous clients.~\cite{wang2019adaptive} demonstrated that FL can be applied on resource-constrained MEC systems to utilize the limited computation and communication resources at the edge efficiently.~\cite{he2020group} propose FedGKT algorithm to reduce the computation overhead in the resource-constrained edge devices.

\textbf{Natural Language Processing.} Gboard, a virtual keyboard from Google for personal gadgets, is one of the notable instances of application of FL in the NLP space. FL helps Gboard learn new words and phrases, thus improving the quality of keyboard search suggestions. There have also been other noteworthy works concerning application of FL in keyboard prediction and keyword spotting over the past few years.~\cite{hard2018federated} proposed using the Federated Averaging technique to train the Coupled Input and Forget Gate (CIFG), a variant of LSTM, for mobile keyboard next-word prediction. In comparison to centralized server-based training, the FL-based approach exhibits a higher precision-recall score. Furthermore, FL was also employed to improve the ranking of browser history suggestions.~\cite{hartmann2019federated} adopted FL for training models on user-interaction data in their work. The authors focused on both privacy issues and robustness of the model in their proposed method.

\textbf{Intrusion Detection.} In the past few years, application of FL in network intrusion detection has significantly improved detection accuracy. A detector proposed by~\cite{zhao2020intelligent} based on FL-assisted LSTM is one of the pioneering works in FL-based intrusion detection.~\cite{liu2021blockchain} also proposed a FL-based intrusion detection system with an emphasis on resource utilization reduction of the central server. In addition, this method incorporates blockchain to safeguard the global model against attacks.~\cite{li2020deepfed} suggested DeepFed, a novel federated deep learning approach for detecting cyber attacks in industrial CPSs (Cyber Physical Systems).

\textbf{Finance.} In recent years a number of financial institutions have implemented FL in areas such as risk management, fraud detection, and anti-money laundering. However, the integration of FL in financial applications are still in the early stages of research.~\cite{zheng2021federated} proposed Deep K-tuplet network, a Federated meta-learning-based approach for detecting credit card transaction frauds.~\cite{cheng2021secureboost} proposed a VFL-based privacy-protection enhancing framework in order to conduct privacy-preserving machine learning on client credit score data. 

The application of federated learning is not confined to the fields mentioned above. The recent years have witnessed the emergence of FL-based use cases outside of the typical scenarios. Researchers have attempted to benefit from the use of FL in domains such as augmented reality (AR) and anomaly detection, among others.  It is reasonable to expect that application areas of FL will grow in the near future.

\section{Challenges}

There are still some challenges of implementing FL, which include privacy, security, communication, systems heterogeneity, statistical heterogeneity, and fairness. 

\textbf{Privacy.} Privacy is a growing concern in FL because the model weights or updates may reflect or disclose sensitive information to a third party, even though there is no original data exchange involved as stated by~\cite{mcmahan2017learning}.~\cite{wahab2021federated} stated that it becomes increasingly difficult to distinguish between regular and malicious clients if the number of clients increases in a collaborative learning environment.
 
By using differential privacy,~\cite{truex2019hybrid} inserted noises into the local gradient updates, after which the noisy updates were then transferred to the central server being encrypted by the Paillier cryptosystem.~\cite{bonawitz2016practical} applied secure multi-party computation  (SMC), a lossless Cryptographic protocol to protect the local individual model updates based on federated averaging. The central server in SMC protocols is only able to observe the final aggregated results, thus protecting local updates.

\textbf{Security.} There are some security concerns in an FL environment as there might be chances of model poisoning and data poisoning attacks. A participant causes model poisoning by attempting to poison the aggregated global model at the server. Consequently, one single attacker has the possibility of poisoning the entire global model. In contrast, synthetic malicious data, also known as dirty-label data, is generated by data poisoning attacks, to adversely contaminate the global model during its training. However, data poisoning attacks are comparatively less harmful since the participant’s updates vary by the number of participants and the dataset.~\cite{bhagoji2019analyzing} annotated the qualitative and quantitative differences between data poisoning and model poisoning and also gave suggestions on model attack prevention.

\textbf{Communication.} In an FL environment,  slow and unstable communication becomes a bottleneck. Since a massive number of heterogeneous devices are involved in training, presumably communication speed gets greatly diminished and packet loss and transmission issues can prevail. Some parties may get disconnected due to unstable wireless communication channels. Hence it is imperative to develop a communication-efficient technique that can either (i) decrease the number of clients, (ii) decrease the number of update rounds, or (iii) decrease transferred message size.

Several communication-efficient learning methods exist to implement these tasks, which can group into three categories (1) decentralized training, (2) local updating methods, and (3) model Compression. Different model compression techniques can decrease message size exchanged at each round such as subsampling, quantization, and sparsification.~\cite{jhunjhunwala2021adaptive} used quantization and random structured rotations to cause the updating models to be sparse.~\cite{caldas2018expanding} employed dropout and lossy compression to lessen device-server communication rounds and reduce client resource requirements.~\cite{sattler2019robust}  adopted ternarization and Golomb encoding of the weight updates.

\textbf{Systems Heterogeneity.} In federated settings, if models vary in terms of resource constraints, unprecedented challenges may arise. Devices may differ in terms of battery power, network connectivity, hardware, and willingness to participate. To handle systems heterogeneity, three key directions can be followed such as (i) asynchronous communication, ii) fault tolerance, or (iii) active device sampling.

\textbf{Statistical Heterogeneity.} Data might not always be identically distributed across devices causing complications to arise. If the training data is not distributed in a balanced manner across the clients, \emph{i.e.}, if the data persists in a non-i.i.d.fashion, it immensely jeopardizes the convergence behavior.~\cite{luo2022tackling} used a mechanism called adaptive client sampling that aims to handle both statistical and systems heterogeneity, diminishing convergence time.~\cite{khodak2019adaptive} used Adaptive Gradient-Based Meta-Learning Methods and have surpassed improved the performance of vanilla FedAvg. Another approach suited to smaller federated settings would be~\cite{corinzia2019variational}, where the clients are treated as a star-shaped Bayesian network and variational inference are performed during learning, which adopts multi-task learning paradigm to reduce strong statistical heterogeneity among real-world FL datasets. Despite approaches and advances toward handling the challenges of FL, some issues remain unsolved in devising automated, scalable, and robust techniques for the heterogeneous settings.

\textbf{Fairness.} The majority of existing FL frameworks or models have not significantly emphasized the issue of fairness in machine learning. Different notions of fairness are emerging in response to the continuously evolving challenges associated with federated learning. Some of the ideas of fairness in FL model training addressed by researchers in recent years include collaborative fairness~\cite{lyu2020collaborative}, which targets providing model updates according to participant contributions, and group fairness~\cite{zhang2020fairfl} for mitigating bias against certain groups of the population. However, this is a relatively new research area, and more successful contributions can be expected from the scientific community in the years to come.

\section{Future Directions}

In this section, we summarize promising directions revolving around the challenges discussed previously such as communication overhead, systems and statistical heterogeneity, privacy and security issues, and fairness ~\cite{li2020federated}.

\textbf{Advanced Communication Schemes.} A direction yet to explore would be to find effective means to compute the exact amount of communication essential in a particular federated system. Errors in optimization methods for machine learning help with generalization as they have some tolerance for lack of precision. FedSVRG by ~\cite{chen2021fl} stands for federated stochastic variance reduced gradient, with an aim to reduce iteration number between the different parties and to maintain the accuracy. However, such methods, which are well explored for conventional settings such as divide-and-conquer and one-shot methods, are not common or understood well in federated or distributed environments.

\textbf{Heterogeneity Diagnostics.} At present, diagnosis of statistical heterogeneity has been made possible before training federated learning models through metrics like earth mover’s distance ~\cite{tong2021diffusion}. These techniques however, render useless prior to the training procedure and hence, it is yet to find out the following things:
\begin{itemize}
    \item  Whether simple diagnostics can quickly determine heterogeneity level prior to the training procedure.

    \item  Whether novel definitions can be developed or existing definitions of heterogeneity be used to ameliorate the convergence of federated optimization techniques.

    \item Whether the development of analogous techniques, capable of qualifying systems heterogeneity is possible. 
\end{itemize}

\textbf{Beyond Traditional Learning.} Data generated in practical federated systems are highly likely to be loosely labeled or unlabeled.  Real-world problems might include conducting complex and sophisticated tasks, such as reinforcement learning. For executing these, exploratory data analysis and aggregate statistics are needed. Complications beyond traditional learning i.e. supervised learning, will needs urgent addressing comparable issues of heterogeneity, privacy and scalability.

\textbf{Benchmarks.} FL, being an emerging field, opens up novel opportunities for shaping the benchmark advancements made in this field so that they could be well incorporated into real-world scenarios and applications. Some open source Systems would be Google TensorFlow Federated (TFF), Federated AI Technology Enabler (FATE)  ~\cite{Cai_2022}, and PySyft ~\cite{ryffel2018generic}, etc. 

One remarkable library that provides baseline implementations for several FL algorithms such as FedAvg, Vertical FL, and split learning and ML models would be FedML ~\cite{he2020fedml}. OARF ~\cite{hu2022oarf}, UniFed ~\cite{liu2022unifed},  Edge AIBench ~\cite{hao2018edge} also provides a diverse set of reference implementations and tools for Federated Learning benchmarks. To allow proper reproducibility and propagation of new solutions, it is pivotal for FL partitioners to build upon the current benchmarking tools such as TensorFlow Federated and LEAF (a moodular benchmarking framework for FL settings) ~\cite{caldas2018leaf}. 

\textbf{Productionization.} Productionizing federated models or deploying these models, in reality, cause problems when new and unrecognized devices make an entree into the network or if they change over some time (concept drift), or when each device demonstrate varying or different behavioral patterns at different hours of a single day (diurnal variation). Robust federated models should not be susceptible to these changes and therefore, appropriate techniques should be introduced that could make federated data generation models immune to these problems.

\textbf{New Methods for Handling Asynchronous Tasks.} Generally, bulk synchronous and asynchronous techniques are prevalent in realistic device-centric communication schemes, where one can become active and communicate with the central server. In these schemes, the workers pull their upcoming tasks from the central code and readily push the outcomes of their previous tasks as opposed to federated settings where the workers remain idle. So it is of paramount importance to come up with mechanisms that can help imbue this characteristic in federated learning environments.

\textbf{Privacy Concerns at Granular Level.} Even though there are existing definitions of privacy that are applicable on a local level or on a global level in a federated network, including all the devices, it is compulsory to introduce methods for granular privacy definition so that device-specific and sample-specific privacy constraints can be handled. This is because privacy limitations vary across different data points on one device as well as across many devices.

\section{Conclusion}

This paper presents an overview of recent research on FL as well as insights from earlier studies. We have also made an effort to provide a summary on the advantages, challenges, and issues related to the design and implementation of FL systems.   Furthermore, several present applications areas of FL have been discussed. Our work is intended to aid researchers and practitioners in the field of artificial intelligence in evaluating the current state of FL research. FL, suggested as an alternative of the centralized machine learning approaches for secure handling of distributed data, has received a lot attention from researchers and academics in recent years. FL is already widely employed as a privacy-preserving solution in several domains. However, there still exist issues and challenges concerning the adoption of FL that need to be addressed in the future. We hope that significant efforts will be invested to overcome these challenges in the future and our paper will serve as an important guideline in defining the research problems related to FL.



\bibliographystyle{named}
\bibliography{ijcai23}

\begin{thebibliography}{}

\bibitem[\protect\citeauthoryear{AbdulRahman~\textit{et
  al}.}{2020}]{abdulrahman2020survey}
Sawsan AbdulRahman~\textit{et al}.
\newblock A survey on federated learning: The journey from centralized to
  distributed on-site learning and beyond.
\newblock {\em IEEE Internet of Things Journal}, 8(7):5476--5497, 2020.

\bibitem[\protect\citeauthoryear{Aledhari~\textit{et
  al}.}{2020}]{aledhari2020federated}
Mohammed Aledhari~\textit{et al}.
\newblock Federated learning: A survey on enabling technologies, protocols, and
  applications.
\newblock {\em IEEE Access}, 8:140699--140725, 2020.

\bibitem[\protect\citeauthoryear{Ammad-Ud-Din~\textit{et
  al}.}{2019}]{ammad2019federated}
Muhammad Ammad-Ud-Din~\textit{et al}.
\newblock Federated collaborative filtering for privacy-preserving personalized
  recommendation system.
\newblock {\em arXiv:1901.09888}, 2019.

\bibitem[\protect\citeauthoryear{Bergou~\textit{et
  al}.}{2022}]{bergou2022personalized}
El~Houcine Bergou~\textit{et al}.
\newblock Personalized federated learning with communication compression.
\newblock {\em arXiv:2209.05148}, 2022.

\bibitem[\protect\citeauthoryear{Bhagoji~\textit{et
  al}.}{2019}]{bhagoji2019analyzing}
Arjun~Nitin Bhagoji~\textit{et al}.
\newblock Analyzing federated learning through an adversarial lens.
\newblock In {\em ICML}, pages 634--643, 2019.

\bibitem[\protect\citeauthoryear{Bonawitz~\textit{et
  al}.}{2016}]{bonawitz2016practical}
Keith Bonawitz~\textit{et al}.
\newblock Practical secure aggregation for federated learning on user-held
  data.
\newblock {\em arXiv:1611.04482}, 2016.

\bibitem[\protect\citeauthoryear{Caldas~\textit{et al}.}{2018}]{caldas2018leaf}
Sebastian Caldas~\textit{et al}.
\newblock Leaf: A benchmark for federated settings.
\newblock {\em arXiv:1812.01097}, 2018.

\bibitem[\protect\citeauthoryear{Caldas\textit{et
  al}.}{2018}]{caldas2018expanding}
Sebastian Caldas\textit{et al}.
\newblock Expanding the reach of federated learning by reducing client resource
  requirements.
\newblock {\em arXiv:1812.07210}, 2018.

\bibitem[\protect\citeauthoryear{Chen and Guestrin}{2016}]{chen2016xgboost}
Tianqi Chen and Carlos Guestrin.
\newblock Xgboost: A scalable tree boosting system.
\newblock In {\em KDD}, pages 785--794, 2016.

\bibitem[\protect\citeauthoryear{Chen~\textit{et
  al}.}{2020a}]{chen2020federated}
Dawei Chen~\textit{et al}.
\newblock Federated learning based mobile edge computing for augmented reality
  applications.
\newblock In {\em ICNC}, pages 767--773, 2020.

\bibitem[\protect\citeauthoryear{Chen~\textit{et al}.}{2020b}]{chen2020joint}
Mingzhe Chen~\textit{et al}.
\newblock A joint learning and communications framework for federated learning
  over wireless networks.
\newblock {\em IEEE Transactions on Wireless Communications}, 20(1):269--283,
  2020.

\bibitem[\protect\citeauthoryear{Chen~\textit{et al}.}{2021}]{chen2021fl}
Shaoqi Chen~\textit{et al}.
\newblock Fl-qsar: a federated learning-based qsar prototype for collaborative
  drug discovery.
\newblock {\em Bioinformatics}, 36(22-23):5492--5498, 2021.

\bibitem[\protect\citeauthoryear{Cheng~\textit{et
  al}.}{2021}]{cheng2021secureboost}
Kewei Cheng~\textit{et al}.
\newblock Secureboost: A lossless federated learning framework.
\newblock {\em IEEE Intelligent Systems}, 36(6):87--98, 2021.

\bibitem[\protect\citeauthoryear{Collins~\textit{et
  al}.}{2021}]{collins2021exploiting}
Liam Collins~\textit{et al}.
\newblock Exploiting shared representations for personalized federated
  learning.
\newblock In {\em ICML}, pages 2089--2099, 2021.

\bibitem[\protect\citeauthoryear{Corinzia \bgroup \em et al.\egroup
  }{2019}]{corinzia2019variational}
Luca Corinzia, Ami Beuret, and Joachim~M Buhmann.
\newblock Variational federated multi-task learning.
\newblock {\em arXiv:1906.06268}, 2019.

\bibitem[\protect\citeauthoryear{Dayan~\textit{et
  al}.}{2021}]{dayan2021federated}
Ittai Dayan~\textit{et al}.
\newblock Federated learning for predicting clinical outcomes in patients with
  covid-19.
\newblock {\em Nature medicine}, 27(10):1735--1743, 2021.

\bibitem[\protect\citeauthoryear{Dongqi~\textit{et al}.}{2022}]{Cai_2022}
Cai Dongqi~\textit{et al}.
\newblock Accelerating vertical federated learning.
\newblock {\em {IEEE} Transactions on Big Data}, pages 1--10, 2022.

\bibitem[\protect\citeauthoryear{Hao~\textit{et al}.}{2018}]{hao2018edge}
Tianshu Hao~\textit{et al}.
\newblock Edge aibench: towards comprehensive end-to-end edge computing
  benchmarking.
\newblock In {\em Bench}, pages 23--30, 2018.

\bibitem[\protect\citeauthoryear{Hard~\textit{et
  al}.}{2018}]{hard2018federated}
Andrew Hard~\textit{et al}.
\newblock Federated learning for mobile keyboard prediction.
\newblock {\em arXiv:1811.03604}, 2018.

\bibitem[\protect\citeauthoryear{Hardy~\textit{et
  al}.}{2017}]{hardy2017private}
Stephen Hardy~\textit{et al}.
\newblock Private federated learning on vertically partitioned data via entity
  resolution and additively homomorphic encryption.
\newblock {\em arXiv:1711.10677}, 2017.

\bibitem[\protect\citeauthoryear{Hartmann~\textit{et
  al}.}{2019}]{hartmann2019federated}
Florian Hartmann~\textit{et al}.
\newblock Federated learning for ranking browser history suggestions.
\newblock {\em arXiv:1911.11807}, 2019.

\bibitem[\protect\citeauthoryear{He \bgroup \em et al.\egroup
  }{2020}]{he2020group}
Chaoyang He, Murali Annavaram, and Salman Avestimehr.
\newblock Group knowledge transfer: Federated learning of large cnns at the
  edge.
\newblock {\em NIPS}, 33:14068--14080, 2020.

\bibitem[\protect\citeauthoryear{He~\textit{et al}.}{2020}]{he2020fedml}
Chaoyang He~\textit{et al}.
\newblock Fedml: A research library and benchmark for federated machine
  learning.
\newblock {\em arXiv:2007.13518}, 2020.

\bibitem[\protect\citeauthoryear{Hu~\textit{et al}.}{2022}]{hu2022oarf}
Sixu Hu~\textit{et al}.
\newblock The oarf benchmark suite: Characterization and implications for
  federated learning systems.
\newblock {\em ACM Transactions on Intelligent Systems and Technology (TIST)},
  13(4):1--32, 2022.

\bibitem[\protect\citeauthoryear{Huang~\textit{et
  al}.}{2019}]{huang2019patient}
Li~Huang~\textit{et al}.
\newblock Patient clustering improves efficiency of federated machine learning
  to predict mortality and hospital stay time using distributed electronic
  medical records.
\newblock {\em Journal of biomedical informatics}, 99:103291, 2019.

\bibitem[\protect\citeauthoryear{Imteaj~\textit{et
  al}.}{2023}]{imteaj2023federated}
Ahmed Imteaj~\textit{et al}.
\newblock Federated learning for resource-constrained iot devices: Panoramas
  and state of the art.
\newblock {\em Federated and Transfer Learning}, pages 7--27, 2023.

\bibitem[\protect\citeauthoryear{Jhunjhunwala~\textit{et
  al}.}{2021}]{jhunjhunwala2021adaptive}
Divyansh Jhunjhunwala~\textit{et al}.
\newblock Adaptive quantization of model updates for communication-efficient
  federated learning.
\newblock In {\em ICASSP}, pages 3110--3114, 2021.

\bibitem[\protect\citeauthoryear{Kairouz~\textit{et
  al}.}{2021}]{kairouz2021advances}
Peter Kairouz~\textit{et al}.
\newblock Advances and open problems in federated learning.
\newblock {\em Foundations and Trends{\textregistered} in Machine Learning},
  14(1--2):1--210, 2021.

\bibitem[\protect\citeauthoryear{Khodak \bgroup \em et al.\egroup
  }{2019}]{khodak2019adaptive}
Mikhail Khodak, Maria-Florina~F Balcan, and Ameet~S Talwalkar.
\newblock Adaptive gradient-based meta-learning methods.
\newblock {\em NIPS}, 32, 2019.

\bibitem[\protect\citeauthoryear{Kone{\v{c}}n{\`y}~\textit{et
  al}.}{2016}]{konevcny2016federated}
Jakub Kone{\v{c}}n{\`y}~\textit{et al}.
\newblock Federated optimization: Distributed machine learning for on-device
  intelligence.
\newblock {\em arXiv:1610.02527}, 2016.

\bibitem[\protect\citeauthoryear{Li~\textit{et al}.}{2019a}]{li2019fair}
Tian Li~\textit{et al}.
\newblock Fair resource allocation in federated learning.
\newblock {\em arXiv:1905.10497}, 2019.

\bibitem[\protect\citeauthoryear{Li~\textit{et al}.}{2019b}]{li2019privacy}
Wenqi Li~\textit{et al}.
\newblock Privacy-preserving federated brain tumour segmentation.
\newblock In {\em MLMI}, pages 133--141, 2019.

\bibitem[\protect\citeauthoryear{Li~\textit{et al}.}{2020a}]{li2020deepfed}
Beibei Li~\textit{et al}.
\newblock Deepfed: Federated deep learning for intrusion detection in
  industrial cyber--physical systems.
\newblock {\em IEEE Transactions on Industrial Informatics}, 17(8):5615--5624,
  2020.

\bibitem[\protect\citeauthoryear{Li~\textit{et al}.}{2020b}]{li2020federated}
Tian Li~\textit{et al}.
\newblock Federated optimization in heterogeneous networks.
\newblock {\em MLSys}, 2:429--450, 2020.

\bibitem[\protect\citeauthoryear{Li~\textit{et al}.}{2020c}]{li2020multi}
Xiaoxiao Li~\textit{et al}.
\newblock Multi-site fmri analysis using privacy-preserving federated learning
  and domain adaptation: Abide results.
\newblock {\em Medical Image Analysis}, 65:101765, 2020.

\bibitem[\protect\citeauthoryear{Li~\textit{et al}.}{2021}]{li2021survey}
Qinbin Li~\textit{et al}.
\newblock A survey on federated learning systems: vision, hype and reality for
  data privacy and protection.
\newblock {\em IEEE Transactions on Knowledge and Data Engineering}, 2021.

\bibitem[\protect\citeauthoryear{Lim~\textit{et al}.}{2020}]{lim2020federated}
Wei Yang~Bryan Lim~\textit{et al}.
\newblock Federated learning in mobile edge networks: A comprehensive survey.
\newblock {\em IEEE Communications Surveys \& Tutorials}, 22(3):2031--2063,
  2020.

\bibitem[\protect\citeauthoryear{Lin~\textit{et al}.}{2020}]{lin2020ensemble}
Tao Lin~\textit{et al}.
\newblock Ensemble distillation for robust model fusion in federated learning.
\newblock {\em NIPS}, 33:2351--2363, 2020.

\bibitem[\protect\citeauthoryear{Liu~\textit{et al}.}{2015}]{liu2015survey}
Ji~Liu~\textit{et al}.
\newblock A survey of data-intensive scientific workflow management.
\newblock {\em Journal of Grid Computing}, 13(4):457--493, 2015.

\bibitem[\protect\citeauthoryear{Liu~\textit{et al}.}{2019}]{liu2019boosting}
Yang Liu~\textit{et al}.
\newblock Boosting privately: Privacy-preserving federated extreme boosting for
  mobile crowdsensing.
\newblock {\em arXiv:1907.10218}, 2019.

\bibitem[\protect\citeauthoryear{Liu~\textit{et al}.}{2021}]{liu2021blockchain}
Hong Liu~\textit{et al}.
\newblock Blockchain and federated learning for collaborative intrusion
  detection in vehicular edge computing.
\newblock {\em IEEE Transactions on Vehicular Technology}, 70(6):6073--6084,
  2021.

\bibitem[\protect\citeauthoryear{Liu~\textit{et al}.}{2022}]{liu2022unifed}
Xiaoyuan Liu~\textit{et al}.
\newblock Unifed: A benchmark for federated learning frameworks.
\newblock {\em arXiv:2207.10308}, 2022.

\bibitem[\protect\citeauthoryear{Luo~\textit{et al}.}{2022}]{luo2022tackling}
Bing Luo~\textit{et al}.
\newblock Tackling system and statistical heterogeneity for federated learning
  with adaptive client sampling.
\newblock In {\em INFOCOM}, pages 1739--1748, 2022.

\bibitem[\protect\citeauthoryear{Lyu \bgroup \em et al.\egroup
  }{2020}]{lyu2020threats}
Lingjuan Lyu, Han Yu, and Qiang Yang.
\newblock Threats to federated learning: A survey.
\newblock {\em arXiv:2003.02133}, 2020.

\bibitem[\protect\citeauthoryear{Lyu~\textit{et
  al}.}{2020}]{lyu2020collaborative}
Lingjuan Lyu~\textit{et al}.
\newblock Collaborative fairness in federated learning.
\newblock In {\em Federated Learning}, pages 189--204. 2020.

\bibitem[\protect\citeauthoryear{Makhija~\textit{et
  al}.}{2022}]{makhija2022architecture}
Disha Makhija~\textit{et al}.
\newblock Architecture agnostic federated learning for neural networks.
\newblock {\em arXiv:2202.07757}, 2022.

\bibitem[\protect\citeauthoryear{Marfoq~\textit{et
  al}.}{2020}]{marfoq2020throughput}
Othmane Marfoq~\textit{et al}.
\newblock Throughput-optimal topology design for cross-silo federated learning.
\newblock {\em NIPS}, 33:19478--19487, 2020.

\bibitem[\protect\citeauthoryear{McMahan~\textit{et
  al}.}{2017a}]{brendan2016communication}
Brendan McMahan~\textit{et al}.
\newblock Communication-efficient learning of deep networks from decentralized
  data.
\newblock In {\em Artificial intelligence and statistics}, pages 1273--1282,
  2017.

\bibitem[\protect\citeauthoryear{McMahan~\textit{et
  al}.}{2017b}]{mcmahan2017learning}
H~Brendan McMahan~\textit{et al}.
\newblock Learning differentially private recurrent language models.
\newblock {\em arXiv:1710.06963}, 2017.

\bibitem[\protect\citeauthoryear{Mohri \bgroup \em et al.\egroup
  }{2019}]{mohri2019agnostic}
Mehryar Mohri, Gary Sivek, and Ananda~Theertha Suresh.
\newblock Agnostic federated learning.
\newblock In {\em ICML}, pages 4615--4625, 2019.

\bibitem[\protect\citeauthoryear{Nikolaenko~\textit{et
  al}.}{2013}]{nikolaenko2013privacy}
Valeria Nikolaenko~\textit{et al}.
\newblock Privacy-preserving ridge regression on hundreds of millions of
  records.
\newblock In {\em IEEE Symposium on Security and Privacy (SP)}, pages 334--348,
  2013.

\bibitem[\protect\citeauthoryear{Nishio and Yonetani}{2019}]{nishio2019client}
Takayuki Nishio and Ryo Yonetani.
\newblock Client selection for federated learning with heterogeneous resources
  in mobile edge.
\newblock In {\em ICC}, pages 1--7, 2019.

\bibitem[\protect\citeauthoryear{Ouyang~\textit{et
  al}.}{2021}]{ouyang2021novel}
Liwei Ouyang~\textit{et al}.
\newblock A novel framework of collaborative early warning for covid-19 based
  on blockchain and smart contracts.
\newblock {\em Information sciences}, 570:124--143, 2021.

\bibitem[\protect\citeauthoryear{Praneeth Karimireddy~\textit{et
  al}.}{2019}]{praneeth2019scaffold}
Sai Praneeth Karimireddy~\textit{et al}.
\newblock Scaffold: Stochastic controlled averaging for federated learning.
\newblock {\em arXiv e-prints}, pages arXiv--1910, 2019.

\bibitem[\protect\citeauthoryear{Ryffel~\textit{et
  al}}{2018}]{ryffel2018generic}
Theo Ryffel~\textit{et al}.
\newblock A generic framework for privacy preserving deep learning.
\newblock {\em arXiv:1811.04017}, 2018.

\bibitem[\protect\citeauthoryear{Sattler~\textit{et
  al}.}{2019}]{sattler2019robust}
Felix Sattler~\textit{et al}.
\newblock Robust and communication-efficient federated learning from non-iid
  data.
\newblock {\em IEEE transactions on neural networks and learning systems},
  31(9):3400--3413, 2019.

\bibitem[\protect\citeauthoryear{Smith~\textit{et
  al}.}{2017}]{smith2017federated}
Virginia Smith~\textit{et al}.
\newblock Federated multi-task learning.
\newblock {\em NIPS}, 30, 2017.

\bibitem[\protect\citeauthoryear{Snyder}{2014}]{yao}
Peter Snyder.
\newblock Yao’s garbled circuits: Recent directions and implementations.
\newblock In {\em Literature Review}. 2014.

\bibitem[\protect\citeauthoryear{Tan~\textit{et al}.}{2020}]{tan2020federated}
Ben Tan~\textit{et al}.
\newblock A federated recommender system for online services.
\newblock In {\em RecSys}, pages 579--581, 2020.

\bibitem[\protect\citeauthoryear{Tong~\textit{et
  al}.}{2021}]{tong2021diffusion}
Alexander~Y Tong~\textit{et al}.
\newblock Diffusion earth mover’s distance and distribution embeddings.
\newblock In {\em ICML}, pages 10336--10346, 2021.

\bibitem[\protect\citeauthoryear{Truex~\textit{et al}.}{2019}]{truex2019hybrid}
Stacey Truex~\textit{et al}.
\newblock A hybrid approach to privacy-preserving federated learning.
\newblock In {\em ACM AIS}, pages 1--11, 2019.

\bibitem[\protect\citeauthoryear{Vepakomma~\textit{et
  al}.}{2018}]{vepakomma2018split}
Praneeth Vepakomma~\textit{et al}.
\newblock Split learning for health: Distributed deep learning without sharing
  raw patient data.
\newblock {\em arXiv:1812.00564}, 2018.

\bibitem[\protect\citeauthoryear{Wahab~\textit{et
  al}.}{2021}]{wahab2021federated}
Omar~Abdel Wahab~\textit{et al}.
\newblock Federated machine learning: Survey, multi-level classification,
  desirable criteria and future directions in communication and networking
  systems.
\newblock {\em IEEE Communications Surveys \& Tutorials}, 23(2):1342--1397,
  2021.

\bibitem[\protect\citeauthoryear{Wang~\textit{et
  al}.}{2019a}]{wang2019adaptive}
Shiqiang Wang~\textit{et al}.
\newblock Adaptive federated learning in resource constrained edge computing
  systems.
\newblock {\em IEEE Journal on Selected Areas in Communications},
  37(6):1205--1221, 2019.

\bibitem[\protect\citeauthoryear{Wang~\textit{et al}.}{2019b}]{wang2019edge}
Xiaofei Wang~\textit{et al}.
\newblock In-edge ai: Intelligentizing mobile edge computing, caching and
  communication by federated learning.
\newblock {\em IEEE Network}, 33(5):156--165, 2019.

\bibitem[\protect\citeauthoryear{Wang~\textit{et al}.}{2019c}]{wang2019beyond}
Zhibo Wang~\textit{et al}.
\newblock Beyond inferring class representatives: User-level privacy leakage
  from federated learning.
\newblock In {\em INFOCOM}, pages 2512--2520, 2019.

\bibitem[\protect\citeauthoryear{Wang~\textit{et
  al}.}{2020}]{wang2020federated}
Hongyi Wang~\textit{et al}.
\newblock Federated learning with matched averaging.
\newblock {\em arXiv:2002.06440}, 2020.

\bibitem[\protect\citeauthoryear{Wang~\textit{et
  al}.}{2021}]{wang2021federated}
Zheng Wang~\textit{et al}.
\newblock Federated learning with fair averaging.
\newblock {\em arXiv:2104.14937}, 2021.

\bibitem[\protect\citeauthoryear{Xiong~\textit{et al}.}{2020}]{xiong2020facing}
Zhaoping Xiong~\textit{et al}.
\newblock Facing small and biased data dilemma in drug discovery with federated
  learning.
\newblock {\em BioRxiv}, 2020.

\bibitem[\protect\citeauthoryear{Yang~\textit{et
  al}.}{2019}]{yang2019federated}
Qiang Yang~\textit{et al}.
\newblock Federated machine learning: Concept and applications.
\newblock {\em ACM Transactions on Intelligent Systems and Technology (TIST)},
  10(2):1--19, 2019.

\bibitem[\protect\citeauthoryear{Yang~\textit{et al}.}{2020}]{yang2020energy}
Zhaohui Yang~\textit{et al}.
\newblock Energy efficient federated learning over wireless communication
  networks.
\newblock {\em IEEE Transactions on Wireless Communications}, 20(3):1935--1949,
  2020.

\bibitem[\protect\citeauthoryear{Yurochkin~\textit{et
  al}.}{2019}]{yurochkin2019bayesian}
Mikhail Yurochkin~\textit{et al}.
\newblock Bayesian nonparametric federated learning of neural networks.
\newblock In {\em ICML}, pages 7252--7261, 2019.

\bibitem[\protect\citeauthoryear{Zhang \bgroup \em et al.\egroup
  }{2020}]{zhang2020fairfl}
Daniel~Yue Zhang, Ziyi Kou, and Dong Wang.
\newblock Fairfl: A fair federated learning approach to reducing demographic
  bias in privacy-sensitive classification models.
\newblock In {\em IEEE Big Data)}, pages 1051--1060, 2020.

\bibitem[\protect\citeauthoryear{Zhang~\textit{et al}.}{2021}]{zhang2021survey}
Chen Zhang~\textit{et al}.
\newblock A survey on federated learning.
\newblock {\em Knowledge-Based Systems}, 216:106775, 2021.

\bibitem[\protect\citeauthoryear{Zhao~\textit{et
  al}.}{2018}]{zhao2018inprivate}
Lingchen Zhao~\textit{et al}.
\newblock Inprivate digging: Enabling tree-based distributed data mining with
  differential privacy.
\newblock In {\em INFOCOM}, pages 2087--2095, 2018.

\bibitem[\protect\citeauthoryear{Zhao~\textit{et
  al}.}{2020}]{zhao2020intelligent}
Ruijie Zhao~\textit{et al}.
\newblock Intelligent intrusion detection based on federated learning aided
  long short-term memory.
\newblock {\em Physical Communication}, 42:101157, 2020.

\bibitem[\protect\citeauthoryear{Zheng~\textit{et
  al}.}{2021}]{zheng2021federated}
Wenbo Zheng~\textit{et al}.
\newblock Federated meta-learning for fraudulent credit card detection.
\newblock In {\em IJCAI}, pages 4654--4660, 2021.

\bibitem[\protect\citeauthoryear{Zhu \bgroup \em et al.\egroup
  }{2021}]{zhu2021federated}
Hangyu Zhu, Haoyu Zhang, and Yaochu Jin.
\newblock From federated learning to federated neural architecture search: a
  survey.
\newblock {\em Complex \& Intelligent Systems}, 7(2):639--657, 2021.

\end{thebibliography}

\end{document}